Kapitza resistance at a domain boundary in linear and nonlinear chains.

Jithu Paul and O.V.Gendelman[*]

Faculty of Mechanical Engineering, Technion – Israel Institute of Technology

Technion City, Haifa, 3200003, Israel

[*] - contacting author, e-mail: ovgend@technion.ac.il

**Abstract**

We explore Kapitza thermal resistance on the boundary between two homogeneous chain fragments with different characteristics. For a linear model, an exact expression for the resistance is derived. In the generic case of frequency mismatch between the domains, the Kapitza resistance is well-defined in the thermodynamic limit. At the same time, in the linear chain, the resistance depends on the thermostat properties and therefore is not a local property of the considered domain boundary. Moreover, if the temperature difference at the ends of the chain is fixed, then neither the temperature drop at the domain boundary nor the heat flux depend on the system size; for the normal transport, one expects the scaling $N^{-1}$ for both. For specific assessment, we consider the case of an isotopic boundary – only the masses in different domains are different. If the domains are nonlinear, but integrable (Toda lattice, elastically colliding particles), the anomalies are similar to the case of linear chain, besides well-articulated thermal dependence of the resistance. For the case of elastically colliding particles, this dependence follows a simple scaling law $R_k \sim T^{-1/2}$. For Fermi-Pasta-Ulam (FPU) domains, both the temperature drop and the heat flux decrease with the chain length, but with different exponents, so the resistance vanishes in the thermodynamic limit. For the domains comprised of rotators, the thermal resistance exhibits the expected normal behavior.

## 1 Introduction

Thermal resistance between two dissimilar materials is a ubiquitous phenomenon, well-known in science and engineering. In the case of a simple one-dimensional heat transport through the boundary, it is usually defined as $R_k = \frac{\Delta T}{J}$, where $\Delta T$ is the temperature drop at the boundary, and $J$ is the heat flux through the boundary. Existence of such a resistance was first mentioned in 1936 [1,2]. Since the first experimental proof given by Kapitza in 1941 [3], numerous experimental and analytical studies were reported [4–7]. The experimental works reported contrasting temperature dependence of Kapitza resistance at very low [8] and high-temperature limits [9]. The first theoretical explanation of the Kapitza resistance is the famous acoustic mismatch model (AMM) [10,11] which is more applicable for classical solids and very low temperature. There were many attempts to improve the AMM [12–16]. At high temperature, diffuse scattering of phonons is the predominant factor responsible for Kapitza resistance [17] and one of the satisfactory theoretical explanation for high temperature experiments is given by the diffuse mismatch model (DMM) [9]. Khalatnikov's theory with some modification for one-dimensional harmonic system is considered in Refs. [18,19]. All known theoretical models are based on certain intermediate scale assumptions, rather than relate the resistance to a particular microstructure. Recent papers [20,21] explored Kapitza resistance in the chain models with isolated defects, and in Ref. [21] the exact solution for the linear chain with linear time-independent inclusion has been derived. As for the nonlinear chain models, in Ref [20], the behavior of the Kapitza resistance in general has been found to correlate with to well-known universality classes with respect to the bulk conductivity [22–28].

In current paper, we generalize the ideas developed in [21] for more complicated and interesting problem of the thermal resistance on the boundary between two chain domains with different properties. For linear systems, the exact analytic solution is derived. As for the nonlinear models, we consider two qualitatively different cases: the chain domains belonging to the class of integrable models (Toda potential, chain of colliding particles) and non-integrable (FPU, chain of rotators). One expects that in realistic system the boundary resistance will be independent on the system size. Below it is demonstrated that in conditions of constant temperature difference between the ends of the chain this situation can be achieved by (at least) two different scenarios – when the thermal drop and the heat flux are both size independent (linear chain, Toda potential, chain of colliding particles), or both scale as $N^{-1}$ in the thermodynamic limit (chain of rotators). To our opinion, only the latter scenario can be considered as "normal", since otherwise the resistance is not a local property of the boundary. The FPU model exhibits somewhat "intermediate" behavior – both the thermal drop and the heat flux scale with the system length, although with different exponents. As a result, the Kapitza resistance between the FPU domains vanishes in the thermodynamic limit.

The paper is structured as follows. In Section 2, the exact analytic solution for the general case of linear domains with a linear time independent inclusion (LTI) is derived. Then, the particular cases of the isotopic domain boundary, and of the perfect frequency matching are explored. In Section 3, Toda lattice and the chain of hard particles with the isotopic boundary are considered. In Section 4 the similar study is performed for the $\beta$-FPU and for the chain of rotators. Section 5 is devoted to the concluding remarks.

## 2 Kapitza resistance at the boundary between the linear domains.

### 2.1 General derivation with the LTI inclusion

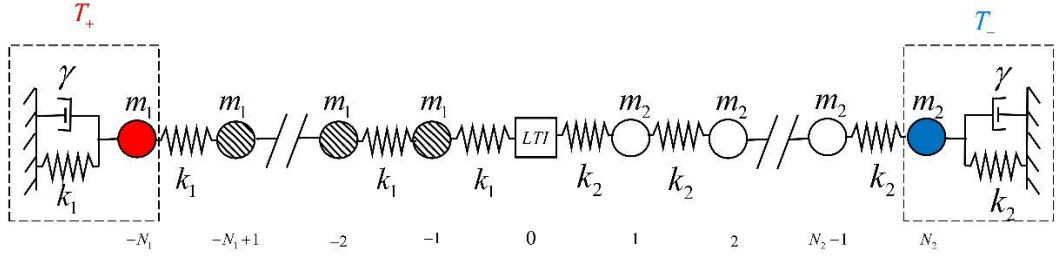

FIG. 1. (Color online) Sketch of the model system

The consideration follows recent paper [21] with necessary modifications. We consider the Kapitza resistance in two sequentially coupled linear chain domains coupled through the LTI inclusion (Fig. 1). The left domain includes $N_1$, and the right domain - $N_2$ particles. In the left domain, the masses and stiffness are set to $m_1$ and $k_1$, and in the right domain - as $m_2$ and $k_2$. The domain boundaries are connected to single Langevin thermostats at both ends. The dynamics of the boundary particles is described by the following equations:

$$m_1 \ddot{u}_{-N_1} + k_1 \left( 2u_{-N_1} - u_{-N_1+1} \right) + \gamma \dot{u}_{-N_1} = \xi_+(t)$$
$$m_2 \ddot{u}_{N_2} + k_2 \left( 2u_{N_2} - u_{N_2-1} \right) + \gamma \dot{u}_{N_2} = \xi_-(t)$$
(1)

Here, $u_n$ represents position of the $n^{th}$ particle. The coupling friction at each end is $\gamma$. $\xi_\pm(t)$ are the white-noise excitations obeying the relation $\langle \xi_\pm(t) \rangle = 0$, $\langle \xi_\pm(t_1)\xi_\pm(t_2) \rangle = 2\gamma T_\pm \delta(t_1 - t_2)$. For the considered linear system, without reducing the generality, the left thermostat is set to temperature $T$ and the right thermostat is set to zero temperature. The particles apart from both the thermostats and the LTI inclusion are described by the equations:

$$m_1 \ddot{u}_n + k_1 (2u_n - u_{n-1} - u_{n+1}) = 0$$
$$m_2 \ddot{u}_n + k_2 (2u_n - u_{n-1} - u_{n+1}) = 0 \qquad (2)$$

The dispersion relations for traveling waves in the chain fragments are presented as

$$\omega_1 = 2\sqrt{\frac{k_1}{m_1}} \sin\left(\frac{q_1}{2}\right) \text{ (Domain 1)}$$
$$\omega_2 = 2\sqrt{\frac{k_2}{m_2}} \sin\left(\frac{q_2}{2}\right) \text{ (Domain 2)} \qquad (3)$$

Here $\omega_i, q_i$, $i=1,2$ denote the frequencies and the wavevectors in the chain fragments. Due to the linearity, in the thermodynamic limit the waves that can transfer energy through the system should belong to the common propagation zone of both domains $\omega \leq \min\left(2\sqrt{\frac{k_1}{m_1}}, 2\sqrt{\frac{k_2}{m_2}}\right)$.

Appropriate conditions for the wave frequencies and the wavevectors are written as follows:

$$\omega = \sqrt{\frac{k_1}{m_1}} \sin\left(\frac{q_1}{2}\right) = \sqrt{\frac{k_2}{m_2}} \sin\left(\frac{q_2}{2}\right)$$
$$-2\arcsin\left(\frac{1}{2}\sqrt{\frac{m_1}{k_1}} \min\left(2\sqrt{\frac{k_1}{m_1}}, 2\sqrt{\frac{k_2}{m_2}}\right)\right) \leq q_1 \leq 2\arcsin\left(\frac{1}{2}\sqrt{\frac{m_1}{k_1}} \min\left(2\sqrt{\frac{k_1}{m_1}}, 2\sqrt{\frac{k_2}{m_2}}\right)\right) \qquad (4)$$
$$-2\arcsin\left(\frac{1}{2}\sqrt{\frac{m_2}{k_2}} \min\left(2\sqrt{\frac{k_1}{m_1}}, 2\sqrt{\frac{k_2}{m_2}}\right)\right) \leq q_2 \leq 2\arcsin\left(\frac{1}{2}\sqrt{\frac{m_2}{k_2}} \min\left(2\sqrt{\frac{k_1}{m_1}}, 2\sqrt{\frac{k_2}{m_2}}\right)\right)$$

The displacements of the particles apart from the boundaries can be presented as Fourier integrals in the common propagation zone (4):

$$u_n = \int_P \exp(i\omega t)(\alpha_1(\omega)\exp(-iq_1 n) + \beta_1(\omega)\exp(iq_1 n))d\omega \text{ (Domain 1)}$$
$$u_n = \int_P \exp(i\omega t)(\alpha_2(\omega)\exp(-iq_2 n) + \beta_2(\omega)\exp(iq_2 n))d\omega \text{ (Domain 2)} \qquad (5)$$

$\alpha_{1,2}$ and $\beta_{1,2}$ are the partial wave amplitudes in the domains. Substituting (5) to (1), one obtains:

$$\alpha_1 \exp(iq_1 N_1)c_{1,+} + \beta_1 \exp(-iq_1 N_1)c_{1,-} = \Xi$$
$$\alpha_2 \exp(-iq_2 N_2)c_{2,-} + \beta_2 \exp(iq_2 N_2)c_{2,+} = 0$$
$$c_{1,\pm} = k_1 \exp(\pm iq_1) + i\gamma\omega$$
$$c_{2,\pm} = k_2 \exp(\pm iq_2) + i\gamma\omega$$

(6)

Here, $\Xi$ represents the Fourier transform of the forcing function $\xi(t)$, and its amplitude can be calculated using Parseval's theorem as $|\Xi|^2 = \dfrac{\gamma T}{\pi}$. The inclusion is linear and time-independent; therefore, it is possible to connect the partial wave amplitudes in the domains by a transfer matrix $\mathbf{G}$:

$$\begin{pmatrix} \alpha_1 \\ \beta_1 \end{pmatrix} = \mathbf{G} \begin{pmatrix} \alpha_2 \\ \beta_2 \end{pmatrix}, \quad \mathbf{G} = \begin{pmatrix} g_{11} & g_{12} \\ g_{21} & g_{22} \end{pmatrix}$$

(7)

Substituting (7) to (6), one obtains:

$$\alpha_2(g_{11}\exp(iq_1 N_1)c_{1,+} + g_{21}\exp(-iq_1 N_1)c_{1,-}) + \beta_2(g_{12}\exp(iq_1 N_1)c_{1,+} + g_{22}\exp(-iq_1 N_1)c_{1,-}) = \Xi$$
$$\alpha_2 \exp(-iq_2 N_2)c_{2,-} + \beta_2 \exp(iq_2 N_2)c_{2,+} = 0$$

(8)

Eq. (8) can be solved for $\alpha_2$ and $\beta_2$:

$$\alpha_2 = \frac{\Xi \exp(iq_2 N_2)c_{2,+}}{D}, \quad \beta_2 = -\frac{\Xi \exp(-iq_2 N_2)c_{2,-}}{D}$$
$$D = g_{11}\exp\left(i(q_1 N_1 + q_2 N_2)\right)c_{1,+}c_{2,+} + g_{21}\exp\left(i(-q_1 N_1 + q_2 N_2)\right)c_{2,+}c_{1,-} -$$
$$- g_{12}\exp\left(i(q_1 N_1 - q_2 N_2)\right)c_{1,+}c_{2,-} - g_{22}\exp\left(-i(q_1 N_1 + q_2 N_2)\right)c_{1,-}c_{2,-}$$

(9)

For finding the most general form of the transfer matrix, first we use the reciprocity theorem [29]. Interchanging the thermostats at the boundaries of the transforms (8) to the following form:

$$\tilde{\alpha}_2(g_{11}\exp(iq_1 N_1)c_{1,+} + g_{21}\exp(-iq_1 N_1)c_{1,-}) + \tilde{\beta}_2(g_{12}\exp(iq_1 N_1)c_{1,+} + g_{22}\exp(-iq_1 N_1)c_{1,-}) = 0$$
$$\tilde{\alpha}_2 \exp(-iq_2 N_2)c_{2,-} + \tilde{\beta}_2 \exp(iq_2 N_2)c_{2,+} = \Xi$$

(10)

$\tilde{\alpha}_{1,2}$ and $\tilde{\beta}_{1,2}$ are the partial wave amplitudes in the chain after interchanging the thermostats.

According to the reciprocity theorem, the displacement in the right domain will be equal to the displacement in the left domain when thermostat positions are interchanged [29]. Then, one obtains

$$\alpha_2 \exp(-iq_2 N_2) + \beta_2 \exp(iq_2 N_2) = \tilde{\alpha}_1 \exp(iq_1 N_1) + \tilde{\beta}_1 \exp(-iq_1 N_1); \quad \begin{pmatrix} \tilde{\alpha}_1 \\ \tilde{\beta}_1 \end{pmatrix} = \mathbf{G} \begin{pmatrix} \tilde{\alpha}_2 \\ \tilde{\beta}_2 \end{pmatrix} \quad (11)$$

This gives the first condition for the elements of the transfer matrix:

$$g_{11} g_{22} - g_{21} g_{12} = \frac{c_{2,+} - c_{2,-}}{c_{1,+} - c_{1,-}} = \frac{\sqrt{k_2 m_2} \cos \frac{q_2}{2}}{\sqrt{k_1 m_1} \cos \frac{q_1}{2}} = \chi \geq 0 \quad (12)$$

We denote the average temperature apart from the thermostats and the boundary, and the heat flux in each domain, as $T_{1,2}$ and $J_{1,2}$ respectively. These parameters can be calculated if one knows the energy density of the waves propagating left wise and right wise in each domain, which is given by $\rho_L(\omega) = \frac{m_i \omega^2}{2} |\alpha_i(\omega)|^2$ and $\rho_R(\omega) = \frac{m_i \omega^2}{2} |\beta_i(\omega)|^2$, $i = 1, 2$. For each domain, the sum of energy densities in each direction yields the temperature and the difference between the energy densities in each direction multiplied by the respective group velocity, $v_{i,gr}(\omega)$, $i = 1, 2$ yields the net heat flux. By taking into consideration the symmetry $\omega \to -\omega$; the expressions can be written as follows:

$$T_1 = \int_P m_1 \omega^2 \left( |\alpha_1|^2 + |\beta_1|^2 \right) d\omega, \quad J_1 = \int_P m_1 \omega^2 \left( |\alpha_1|^2 - |\beta_1|^2 \right) |v_{1,gr}| d\omega;$$

$$T_2 = \int_P m_2 \omega^2 \left( |\alpha_2|^2 + |\beta_2|^2 \right) d\omega, \quad J_2 = \int_P m_2 \omega^2 \left( |\alpha_2|^2 - |\beta_2|^2 \right) |v_{2,gr}| d\omega; \quad (13)$$

$$|v_{1,gr}| = \left| \frac{d\omega}{dq_1} \right| = \left| \sqrt{\frac{k_1}{m_1}} \cos \frac{q_1}{2} \right|, \quad |v_{2,gr}| = \left| \frac{d\omega}{dq_2} \right| = \left| \sqrt{\frac{k_2}{m_2}} \cos \frac{q_2}{2} \right|$$

At this stage, we assume that the inclusion does not contain any dissipative elements, Energy conservation condition $J_1 = J_2$ yields:

$$m_1\left(|\alpha_1|^2-|\beta_1|^2\right)\left|\sqrt{\frac{k_1}{m_1}}\cos\frac{q_1}{2}\right|=m_2\left(|\alpha_2|^2-|\beta_2|^2\right)\left|\sqrt{\frac{k_2}{m_2}}\cos\frac{q_2}{2}\right|$$

(14)

The second set of conditions from the heat flux equality can be calculated as:

$$|g_{11}|^2-|g_{21}|^2=\frac{\left|\sqrt{m_2k_2}\cos\frac{q_2}{2}\right|}{\left|\sqrt{m_1k_1}\cos\frac{q_1}{2}\right|}=\chi;\quad |g_{22}|^2-|g_{12}|^2=\frac{\left|\sqrt{m_2k_2}\cos\frac{q_2}{2}\right|}{\left|\sqrt{m_1k_1}\cos\frac{q_1}{2}\right|}=\chi;\quad g_{11}g_{12}^*=g_{21}g_{22}^*$$

(15)

From (12) and (15), we find that for linear chain fragments, the transfer matrix can be presented in the following general form:

$$\mathbf{G}=\sqrt{\chi}\begin{pmatrix}\cosh x\exp(i\theta_1) & \sinh x\exp(-i\theta_2)\\ \sinh x\exp(i\theta_2) & \cosh x\exp(-i\theta_1)\end{pmatrix};\quad x\in[0,\infty),\theta_1,\theta_2\in[0,2\pi)$$

(16)

Coefficients $\alpha_{1,2}$ and $\beta_{1,2}$ are found by substituting of (16) to (9) and (7). The heat flux through the chain, and average temperatures on each fragment far away from the boundaries are expressed as follows:

$$J=\int_P m_2\omega^2\left(|\alpha_2|^2-|\beta_2|^2\right)|v_{gr}|d\omega=\int_P m_2\omega^2|\Xi|^2|v_{2,gr}|\frac{|c_{2,+}|^2-|c_{2,-}|^2}{|D|^2}d\omega;$$

$$T_1=\int_P m_1\omega^2\left(|\alpha_1|^2+|\beta_1|^2\right)d\omega=$$

$$=\int_P\frac{m_1\chi\omega^2}{|D|^2}|\Xi|^2\begin{bmatrix}(\cosh^2 x+\sinh^2 x)(|c_{2,+}|^2+|c_{2,-}|^2)-\\ -\sinh 2x(c_{2,+}c_{2,-}^*\exp i(\theta_1+\theta_2+2qN)+c_{2,+}^*c_{2,-}\exp i(-\theta_1-\theta_2-2qN))\end{bmatrix}d\omega;$$

$$T_2=\int_P m_2\omega^2\left(|\alpha_2|^2+|\beta_2|^2\right)d\omega=\int_P m_2\omega^2|\Xi|^2\frac{|c_{2,+}|^2+|c_{2,-}|^2}{|D|^2}d\omega;$$

$$D=\sqrt{\chi}\begin{pmatrix}\cosh x\exp i\left((q_1N_1+q_2N_2)+\theta_1\right)c_{1,+}c_{2,+}+\sinh x\left(c_{2,+}c_{1,-}e^{i\theta_2}-c_{1,+}c_{2,-}e^{-i\theta_2}\right)-\\ -\cosh x\exp i\left(-\theta_1-(q_1N_1+q_2N_2)\right)c_{1,-}c_{2,-}\end{pmatrix}$$

(17)

Further step is the transition to thermodynamic limit $N_{1,2} \to \infty$. Setting aside for a moment the case of a perfect acoustical matching between the domains with $\sqrt{k_1/m_1} = \sqrt{k_2/m_2}$, one can note that the relationship between the wavevectors (4) is expressed by the transcendental equation. Therefore, for almost all values of frequency the ratio between the rapidly oscillating phases $\varphi_{1,2} = q_{1,2} N_{1,2}$ will be irrational and the transition to the thermodynamic limit should be performed through a non-resonant averaging:

$$\langle F(\exp(i\varphi_1), \exp(i\varphi_2)) \rangle_{\substack{N_1 \to \infty \\ N_2 \to \infty}} = \frac{1}{4\pi^2} \int_0^{2\pi} d\varphi_1 \int_0^{2\pi} d\varphi_2 F(\exp(i\varphi_1), \exp(i\varphi_2)) \tag{18}$$

The averaging in Eq. (17) yields (see Appendix A):

$$\left\langle \frac{1}{|D|^2} \right\rangle_{\substack{N_1 \to \infty \\ N_2 \to \infty}} = \frac{1}{\chi \sqrt{Q^2 - 4PP^*}} \tag{19}$$

$$\left\langle \frac{\exp(i(2\varphi_2 + \theta_1 + \theta_2))}{|D|^2} \right\rangle_{\substack{N_1 \to \infty \\ N_2 \to \infty}} = \frac{Q - \sqrt{Q^2 - 4PP^*}}{2\chi P \sqrt{Q^2 - 4PP^*}} \tag{20}$$

Here

$$Q = \left(|c_{1,+}|^2 |c_{2,+}|^2 - |c_{1,-}|^2 |c_{2,-}|^2\right) \cosh^2 x + \left(|c_{1,+}|^2 |c_{2,-}|^2 - |c_{2,+}|^2 |c_{1,-}|^2\right) \sinh^2 x$$
$$P = \left(|c_{1,+}|^2 - |c_{1,-}|^2\right) \sinh x \cosh x \left(c_{2,+} c_{2,-}^*\right) \tag{21}$$

Applying (19), (20) and (21) in (17) results in the following set of equations for the non-resonant case.

$$T_1 = \begin{cases} 2\left( \int_0^{\min\left(2\sqrt{\frac{k_1}{m_1}},2\sqrt{\frac{k_2}{m_2}}\right)} \frac{m_1\omega^2}{\sqrt{Q^2-4PP^*}} \frac{\gamma T}{\pi} \left[ \begin{array}{c} (\cosh^2 x + \sinh^2 x)\left(|c_{2,+}|^2 + |c_{2,-}|^2\right) - \\ -\frac{2Q}{\left(|c_{1,+}|^2 - |c_{1,-}|^2\right)} \end{array} \right] d\omega + \\ + \int_0^{\max\left(2\sqrt{\frac{k_1}{m_1}},2\sqrt{\frac{k_2}{m_2}}\right)} \frac{m_1\omega^2 \gamma T}{\pi} \left[ \frac{2}{\left(|c_{1,+}|^2 - |c_{1,-}|^2\right)} \right] d\omega \right), \text{ if } m_1 < 1 \\[2em] 2\left( \int_0^{\min\left(2\sqrt{\frac{k_1}{m_1}},2\sqrt{\frac{k_2}{m_2}}\right)} \frac{m_1\omega^2}{\sqrt{Q^2-4PP^*}} \frac{\gamma T}{\pi} \left[ \begin{array}{c} (\cosh^2 x + \sinh^2 x)\left(|c_{2,+}|^2 + |c_{2,-}|^2\right) - \\ -\frac{2Q}{\left(|c_{1,+}|^2 - |c_{1,-}|^2\right)} \end{array} \right] d\omega + \\ + \int_0^{\min\left(2\sqrt{\frac{k_1}{m_1}},2\sqrt{\frac{k_2}{m_2}}\right)} \frac{m_1\omega^2 \gamma T}{\pi} \left[ \frac{2}{\left(|c_{1,+}|^2 - |c_{1,-}|^2\right)} \right] d\omega \right), \text{ if } m_1 > 1 \end{cases} =$$

$$= T - 2\left( \int_0^{\min\left(2\sqrt{\frac{k_1}{m_1}},2\sqrt{\frac{k_2}{m_2}}\right)} \frac{m_1\omega^2}{\sqrt{Q^2-4PP^*}} \frac{\gamma T}{\pi} \left[ \begin{array}{c} \frac{2Q}{\left(|c_{1,+}|^2 - |c_{1,-}|^2\right)} - \\ -(\cosh^2 x + \sinh^2 x)\left(|c_{2,+}|^2 + |c_{2,-}|^2\right) \end{array} \right] d\omega \right)$$

$$T_2 = 2 \int_0^{\min\left(2\sqrt{\frac{k_1}{m_1}},2\sqrt{\frac{k_2}{m_2}}\right)} m_2\omega^2 \frac{\gamma T}{\pi} \frac{|c_{2,+}|^2 + |c_{2,-}|^2}{\chi\sqrt{Q^2-4PP^*}} d\omega;$$

$$J = 2 \int_0^{\min\left(2\sqrt{\frac{k_1}{m_1}},2\sqrt{\frac{k_2}{m_2}}\right)} m_2\omega^2 \frac{\gamma T}{\pi} |v_{2,gr}| \frac{|c_{2,+}|^2 - |c_{2,-}|^2}{\chi\sqrt{Q^2-4PP^*}} d\omega;$$

(22)

The Kapitza resistance is given by $R_K = \dfrac{T_1 - T_2}{J}$. (23)

### 2.2 Particular system: isotopic interface

Here we consider the case of the isotopic domain, where the LTI is a single unit mass at $n=0$, all stiffness coefficients and masses $m_2$ are set to unity. To find the elements of the transfer matrix, we consider the particle with $n=0$:

$$\ddot{u}_0 + 2u_0 - u_1 - u_{-1} = 0 \tag{24}$$

Substituting (5) to (24), one obtains:

$$\begin{aligned}&\alpha_1 + \beta_1 = \alpha_2 + \beta_2 \\ &\exp(iq_1)\alpha_1 + \exp(-iq_1)\beta_1 = \\ &= (2 - \omega^2 - \exp(-iq_2))\alpha_2 + (2 - \omega^2 - \exp(iq_2))\beta_2\end{aligned} \tag{25}$$

From (25) one obtains the transfer matrix in the following form:

$$\begin{aligned}\mathbf{G} &= \frac{i}{2\sin q_1}\begin{pmatrix} e^{-iq_1} + \omega^2 + e^{-iq_2} - 2 & e^{-iq_1} + \omega^2 + e^{iq_2} - 2 \\ -e^{iq_1} - \omega^2 - e^{-iq_2} + 2 & -e^{iq_1} - \omega^2 - e^{iq_2} + 2 \end{pmatrix} = \\ &= \frac{i}{2\sin q_1}\begin{pmatrix} e^{-iq_1} - e^{iq_2} & e^{-iq_1} - e^{-iq_2} \\ -e^{iq_1} + e^{iq_2} & -e^{iq_1} + e^{-iq_2} \end{pmatrix}\end{aligned} \tag{26}$$

Some further simplification yields:

$$\begin{aligned}\cosh^2 x &= \frac{2(m_1+1) - m_1\omega^2 + \sqrt{(4m_1 - m_1^2\omega^2)(4 - \omega^2)}}{2\chi(4m_1 - m_1^2\omega^2)}; \\ \sinh^2 x &= \frac{2(m_1+1) - m_1\omega^2 - \sqrt{(4m_1 - m_1^2\omega^2)(4 - \omega^2)}}{2\chi(4m_1 - m_1^2\omega^2)}; \\ \chi &= \frac{1}{\sqrt{m_1}}\frac{\sqrt{4-\omega^2}}{\sqrt{4-m_1\omega^2}}\end{aligned} \tag{27}$$

Evaluation of the Kapitza resistance using (22), (23) and (27) demonstrates clear convergence to the average value in the thermodynamic limit (18) for various relative positions of the domain boundary with increasing total length (Fig. 2).

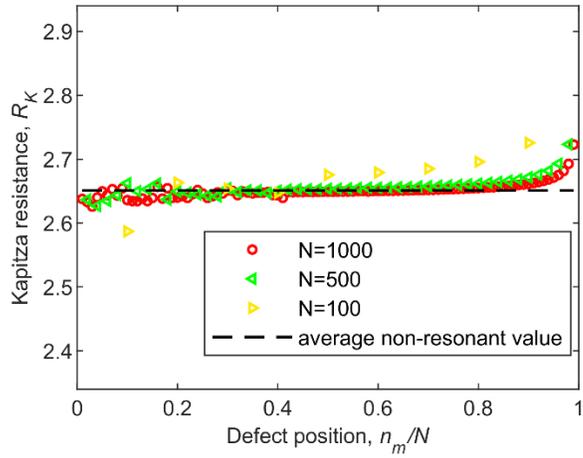

FIG. 2. (Color online) For the isotopic interface problem, the Kapitza resistance is plotted for various interface positions in the chain for various chain lengths using the numerical solution [22]. Horizontal line corresponds to the non-resonant background value obtained from (22, 23), $T=1, m_1=2, \gamma=1$

Dependencies of the resistance on the mismatch $m_1$ and dissipation coefficient in the thermostat $\gamma$ are presented in Figs. 3 and 4.

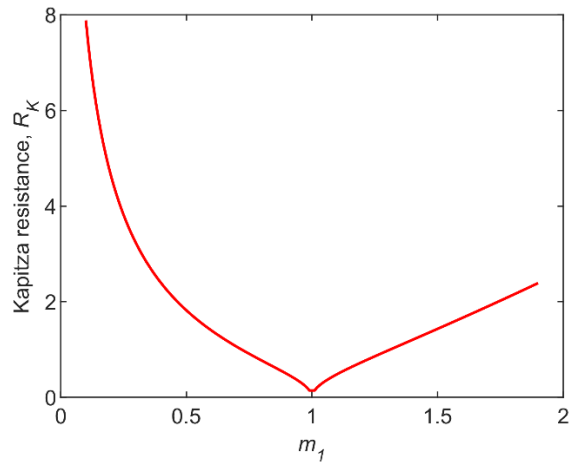

FIG. 3. (Color online) Dependence of Kapitza resistance on mass $m_1$ for the isotopic interface problem. $T=1, \gamma=1$

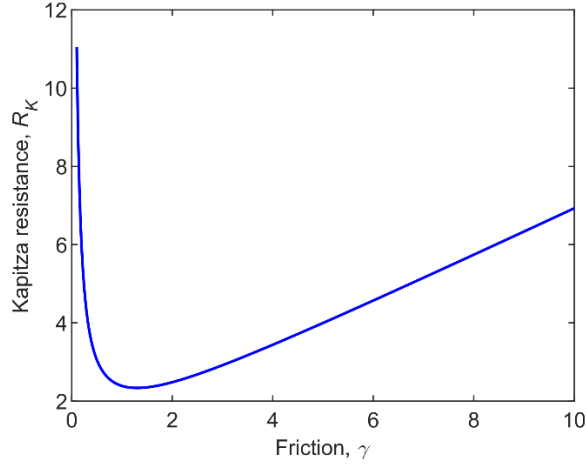

FIG. 4. (Color online) Dependence of Kapitza resistance on the thermostat friction $\gamma$ for the isotopic interface problem. $T=1, m_1 =1.9$

To derive the "cusp" in Figure 3, one sets $\varepsilon =|m_1 -1|\ll 1$, and obtains:

$$\cosh^2 x = \frac{2(2+\varepsilon)-(1+\varepsilon)\omega^2}{2\sqrt{4-\omega^2}\sqrt{4(1+\varepsilon)-(1+\varepsilon)^2\omega^2}} + \frac{1}{2} = 1+O(\varepsilon^2)$$
$$\sinh^2 x = O(\varepsilon^2)$$
(28)

In addition, $\sqrt{Q^2 - 4PP^*} = Q + O(\varepsilon^2)$. Then, using (28) and expanding the terms $|c_{1,\pm}|^2$ and $|c_{2,\pm}|^2$, the asymptotic estimations for $T_1$, $T_2$ and $J$ are obtained as follows:

$$T_1 \approx T - 2\left(\int_0^{\min\left(2\sqrt{\frac{1}{m_1}},2\right)} \frac{m_1\omega^2}{Q} \frac{\gamma T}{\pi}\left[\frac{2Q}{\left(|c_{1,+}|^2 - |c_{1,-}|^2\right)} - \left(|c_{2,+}|^2 + |c_{2,-}|^2\right)\right]d\omega\right) \approx$$

$$\approx \begin{cases} \int_0^{2\sqrt{\frac{1}{m_1}}} 2m_1\omega^2 \frac{\gamma T}{\pi} \frac{|c_{2,+}|^2 + |c_{2,-}|^2}{|c_{1,+}|^2|c_{2,+}|^2 - |c_{1,-}|^2|c_{2,-}|^2} d\omega, \text{ if } m_1 > 1 \\ T - \frac{2\sin^{-1}\left(\sqrt{m_1}\right)T}{\pi} + \int_0^2 2m_1\omega^2 \frac{\gamma T}{\pi} \frac{|c_{2,+}|^2 + |c_{2,-}|^2}{|c_{1,+}|^2|c_{2,+}|^2 - |c_{1,-}|^2|c_{2,-}|^2} d\omega, \text{ if } m_1 < 1 \end{cases} =$$

$$= \begin{cases} \int_0^{2\sqrt{\frac{1}{m_1}}} \frac{2m_1 T}{\pi} \frac{1}{\sqrt{4-\omega^2} + \sqrt{4m_1 - m_1^2\omega^2}} d\omega, \text{ if } m_1 > 1 \\ T - \frac{2\sin^{-1}\left(\sqrt{m_1}\right)T}{\pi} + \int_0^2 \frac{2m_1 T}{\pi} \frac{1}{\sqrt{4-\omega^2} + \sqrt{4m_1 - m_1^2\omega^2}} d\omega, \text{ if } m_1 < 1 \end{cases} =$$

$$= \begin{cases} \frac{T}{2}, \text{ if } m_1 > 1 \\ T - \frac{T}{\pi}\sin^{-1}\left(\sqrt{m_1}\right) \approx T - \frac{T}{\pi}\left(\frac{\pi}{2} - \sqrt{m_1 - 1}\right) \approx \frac{T}{2} + \frac{T}{\pi}\sqrt{m_1 - 1}, \text{ if } m_1 < 1 \end{cases} \quad (29)$$

Similarly,

$$T_2 \approx \begin{cases} \frac{T}{\pi}\sin^{-1}\left(\frac{1}{\sqrt{m_1}}\right) \approx \frac{T}{\pi}\left(\frac{\pi}{2} - \sqrt{m_1 - 1}\right) \approx \frac{T}{2} - \frac{T}{\pi}\sqrt{m_1 - 1}, \text{ if } m_1 > 1 \\ \frac{T}{2}, \text{ if } m_1 < 1 \end{cases} \quad (30)$$

$$J \approx \begin{cases} \int_0^{2\sqrt{\frac{1}{m_1}}} \frac{\gamma T\sqrt{m_1}}{\pi} \frac{\omega^2\left(4-\omega^2\right)}{\left(1+\gamma^2\omega^2\right)\left(\frac{4-\omega^2}{\sqrt{4-m_1\omega^2}} + \sqrt{m_1}\sqrt{4-\omega^2}\right)} d\omega, \text{ if } m_1 > 1 \\ \int_0^2 \frac{\gamma T\sqrt{m_1}}{\pi} \frac{\omega^2\left(4-\omega^2\right)}{\left(1+\gamma^2\omega^2\right)\left(\frac{4-\omega^2}{\sqrt{4-m_1\omega^2}} + \sqrt{m_1}\sqrt{4-\omega^2}\right)} d\omega, \text{ if } m_1 < 1 \end{cases} = \frac{\gamma T}{2} \quad (31)$$

Then the Kapitza resistance exhibits the "cusp" presented in Fig. 3:

$$T_1 - T_2 \approx \frac{T}{\pi}\sqrt{|m_1 - 1|}; J \approx \frac{1}{2}\gamma T; R_K \approx \frac{2\sqrt{|m_1 - 1|}}{\pi\gamma} \qquad (32)$$

For the opposite asymptotic limit of very large mass mismatch one assumes $|m_1 - 1| \gg 1, \gamma \ll 1$, and obtains the following estimation for the heat flux:

$$J \approx \int_0^{\frac{2}{\sqrt{m_1}}} \frac{\gamma T}{\pi}\omega^2\sqrt{4-\omega^2}\frac{4\gamma\omega^2}{\frac{\sqrt{4-\omega^2}}{\sqrt{m_1}\sqrt{4-m_1\omega^2}}4\gamma\omega^2\sqrt{m_1}}d\omega = \int_0^{\frac{2}{\sqrt{m_1}}} \frac{\gamma T}{\pi}\omega^2\sqrt{4-m_1\omega^2}d\omega = \frac{\gamma T}{m_1^{3/2}}$$

$$(33)$$

## 2.3   Particular system: perfect acoustic match

For primarily academic purposes, we treat also the case of the perfect acoustic match between the domains. Accordingly, it is assumed that $\frac{m_1}{k_1} = \frac{m_2}{k_2}$ (accordingly, $q_1 = q_2$) and that the domains are connected through mass $m_2$. Here we do not expect the existence of thermodynamic limit, since the rapidly oscillating phases $\varphi_{1,2}$ can be locked-in, and both non-resonant and resonant cases will appear, similarly to [21]. The dependence of the Kapitza resistance on the interface position is plotted in Fig. 5 and the resonant cases readily reveal themselves on the non-resonant background. Since $q_1 = q_2$, the transfer matrix can be simplified as follows (similar to (24)-(26)):

$$\mathbf{G} = \frac{i}{2k_1\sin q}\begin{pmatrix} k_1e^{-iq} - k_2e^{iq} - k_1 + k_2 & k_1e^{-iq} - k_2e^{-iq} - k_1 + k_2 \\ -k_1e^{iq} + k_2e^{iq} + k_1 - k_2 & -k_1e^{iq} + k_2e^{-iq} + k_1 - k_2 \end{pmatrix} \qquad (34)$$

$$\cosh^2 x = \frac{(k_1-k_2)^2(1-\cos q) + 2k_1k_2\sin^2 q}{2\chi k_1^2\sin^2 q}; \sinh^2 x = \frac{(k_1-k_2)^2(1-\cos q)}{2\chi k_1^2\sin^2 q}; \chi = \frac{k_2}{k_1} \qquad (35)$$

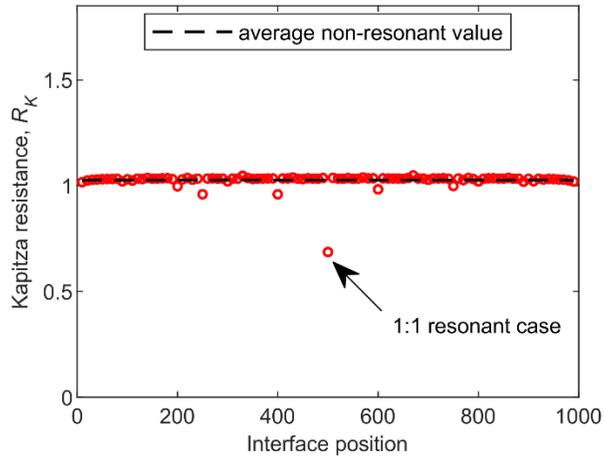

FIG. 5. (Color online) Kapitza resistance for the isotopic boundary in the case of the perfect acoustical match, for various boundary positions in the chain with fixed length. Horizontal line corresponds to the non-resonant background value, $T=1, m_1=1.9, \gamma=1, N_1+N_2=1000$.

Dependences on the resistance on $m_1$ and $\gamma$ are presented in Figs. 6 and 7.

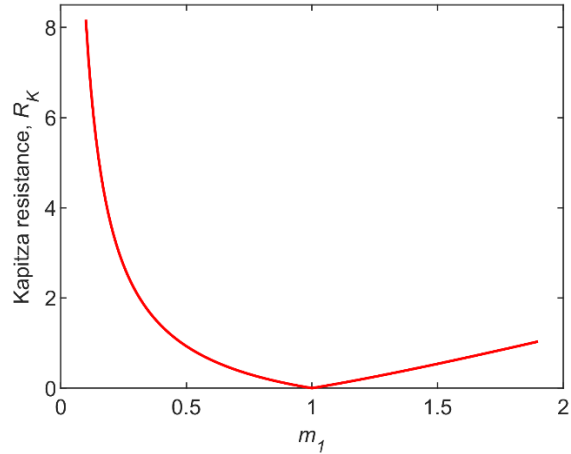

FIG. 6. (Color online) Non-resonant Kapitza resistance at the isotopic boundary in the conditions of perfect acoustical match on $m_1$. Here, $T=1, \gamma=1, m_2=1, \dfrac{m_1}{k_1}=\dfrac{m_2}{k_2}=4$

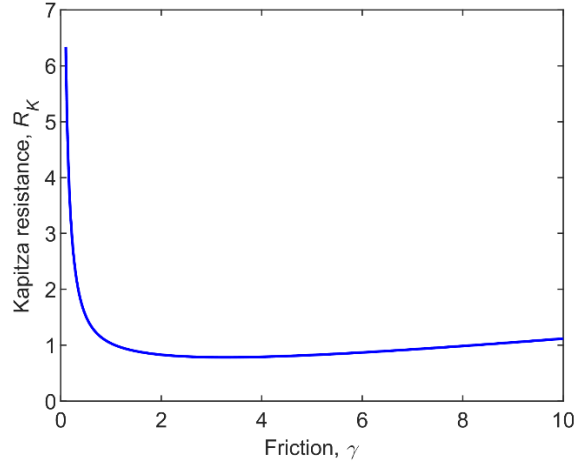

FIG.7. (Color online) Non-resonant Kapitza resistance at the isotopic boundary in the conditions of perfect acoustical match on coupling friction $\gamma$, $T=1, m_1=1.9, m_2=1, \dfrac{m_1}{k_1}=\dfrac{m_2}{k_2}=4$

For the asymptotic case $|m_1-m_2|\ll 1, |k_1-k_2|\ll 1, \gamma\ll 1$, one obtains the following asymptotic limit:

$$R_K \approx \frac{|m_1-m_2|}{\gamma} \qquad (36)$$

The Kapitza resistance again demonstrates non-analytic behavior in the limit of small mass mismatch, but the exponent differs from the generic case of no-matching (32).

### 3    Isotopic boundary between nonlinear integrable domains.

In this section, we consider the isotopic boundary between intergable fragments - Toda lattice and its high-temperature limit, the chain of rigid colliding particles. These models are integrable and have many similarities with the harmonic model such as undefined thermal conductivity since the temperature profile is horizontal. One also expects that the boundary resistance will be well-defined; the substantial difference compared to the linear case is the temperature dependence. From this section onwards, unless mentioned otherwise, the thermostat temperatures are set to $T_\pm = T(1\pm\Delta)$, with $\Delta=0.1$. We use the molecular dynamics simulation with Verlet algorithm with time step ranging $dt=0.001-0.01$. For each simulation, a simulation length of $\tau=10^9$ time steps are used with 640 realizations. The error bar is smaller than the size of the marker.

## 3.1 Toda lattice

The nearest neighbor interaction potential for the Toda lattice [30] is given by,

$$V(u) = \exp(u) \qquad (37)$$

The typical temperature and heat flux profiles in the Toda lattice are shown in Fig. 8; it is clear that local thermal equilibrium is absent for this case and the temperature profile is flat. The Kapitza resistance here is size-independent and depends on the thermostats, similarly to the linear model. First, the asymptotic limit of very high mass mismatch is checked for moderate temperatures $T = 1-4$. As shown in Fig. 9, the temperature and mass dependence are close to Eq. (33).

In Fig. 10, the dependence of Kapitza resistance on the mass mismatch is presented. The Toda lattice in the limit of very low and high temperatures behaves like harmonic model and the chain of colliding particles respectively (the latter is considered in next section). As shown in the inset of Fig. 10, at very low temperatures, one observes $R_K \sim |m_1 - 1|^{0.5}$, similarly to the harmonic model; at moderate temperatures $R_K \sim |m_1 - 1|^{0.7}$ and again to $R_K \sim |m_1 - 1|^{0.5}$ at very high temperature where the Toda lattice model behaves like the colliding particles.

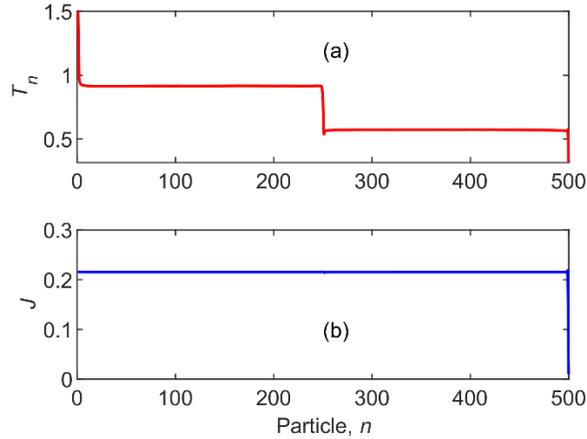

FIG. 8. (Color online) Temperature profile of the Toda lattice model with isotopic-interface is shown in (a) and the heat flux profile is shown in (b). $N = 500, T_+ = 1.9, T_- = 0.1, m_1 = 1.9$

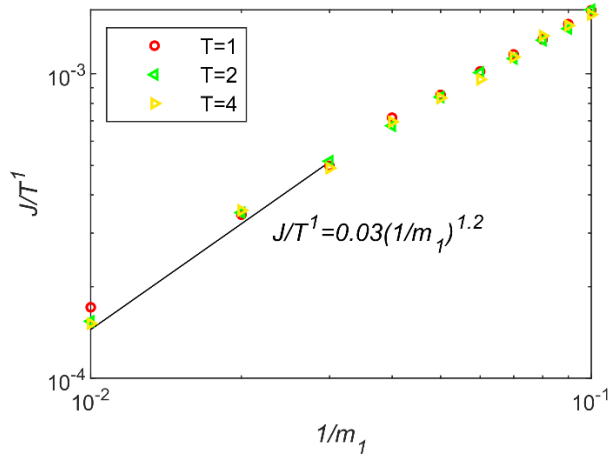

FIG. 9. (Color online) For the isotopic-interface problem, heat flux variation with the chain temperature and $m_1$ for Toda lattice model at $\gamma \ll 1, |m_1 - 1| \gg 1$. Here $N = 500, \gamma = 0.1, \Delta = 0.1$

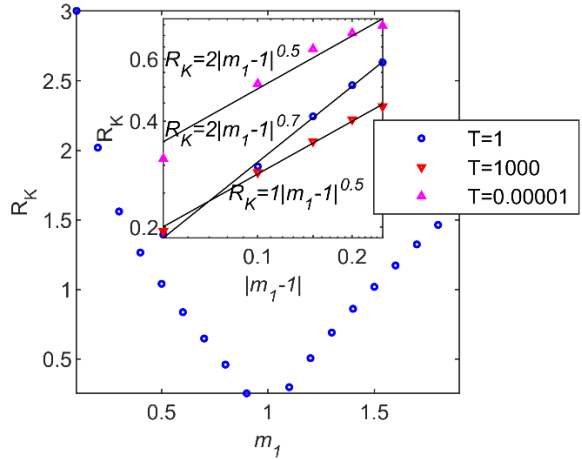

FIG. 10. (Color online) Kapitza resistance versus the mass mismatch in the Toda model. The inset shows asymptotic dependence of Kapitza resistance on small mass mismatch at very small to high temperatures. The cross-over from harmonic case to hard particle case is observed. $N = 500, \Delta = 0.1$

The thermal dependence of the Kapitza resistance is presented in Figure 11. One can observe the crossover from the flat dependence at the low temperatures (the quasilinear case) to (presumably) the power law at high temperatures.

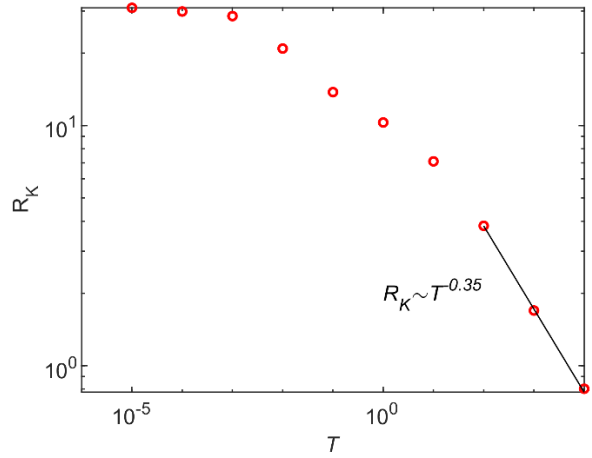

FIG. 11. (Color online) The thermal dependence of Kapitza resistance and the crossover from harmonic case to hard particle case for the Toda model. $N=500, \Delta=0.1$

### 3.2 Chain of colliding particles

Next, let us consider the chain of colliding particles. Separate consideration of this high-temperature limit of the Toda model is justified, since we use an alternative (Maxwell) thermostat. The setting of the model is as follows: $N$ number of hard particles in a chain of length $L$ with a lattice constant $A$ ($A=1$) is considered. The two boundaries of the system are connected to the Maxwell wall at different temperatures $T_+$ and $T_-$. The simulation is based on event-driven algorithm, total simulation length of $10^9$ events is considered. The velocities of each colliding pair after the collision is calculated using the energy and momentum conservation laws as follows,

$$v_i' = \frac{m_i - m_{i+1}}{m_i + m_{i+1}} v_i + \frac{2m_{i+1}}{m_i + m_{i+1}} v_{i+1}$$

$$v_{i+1}' = \frac{2m_i}{m_i + m_{i+1}} v_i - \frac{m_i - m_{i+1}}{m_i + m_{i+1}} v_{i+1} \qquad (38)$$

Here, $v$ is the velocity and the prime symbol represents the updated velocities. When the particle hits the wall, it is reflected at the velocity according to the Maxwell distribution:

$$P(v) = m|v|/T_\pm exp\left[-mv^2/(2T_\pm)\right] \qquad (39)$$

In Fig. 12, the Kapitza resistance dependence on the left fragment mass is shown. As it is clear from the inset of Fig. 12, the trend is very similar to the harmonic case and high-temperature case of Toda lattice, i.e. Kapitza resistance obeys square root dependence on the mass mismatch at $|m_1 - 1| \ll 1$.

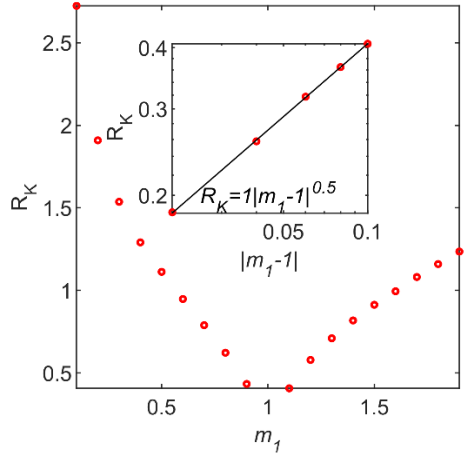

FIG. 12. (Color online) Kapitza resistance for the chain of hard particles model versus the mass mismatch, $N = 500, T_+ = 1.9, T_- = 0.1$). The inset shows the asymptotic dependence of Kapitza resistance on small mass mismatch.

The temperature dependence of the chain of colliding particles is easy to estimate. The Kapitza resistance is defined as $R_K = \dfrac{\Delta T}{J}$. Here, $\Delta T \sim T$, since in the chain of colliding particles, there are no additional parameters with the dimensionality of energy and the heat flux is proportional to $T$ multiplied by the particle velocity, i.e. $J \sim T^{3/2}$, therefore Kapitza resistance behaves like $R_K \sim T^{-1/2}$. One can conjecture that the exponent denoted in Figure 11 is not final, and for even higher temperatures one should obtain the square root dependence.

For $|m_1 - 1| \gg 1$, the heat flux dependence on the mass $m_1$ is shown in Fig. 13. As it is derived in the previous paragraph, the heat flux dependence on temperature is $J \sim T^{3/2}$ (this is exact exponent) and the mass dependence is as given by the asymptotic Eq. (33).

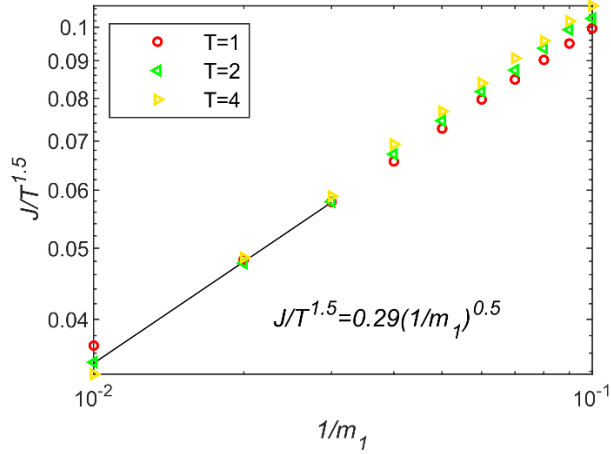

FIG. 13. (Color online) Heat flux variation with the chain temperature and $m_1$ for chain of colliding particle as $|m_1 - 1| \gg 1$. Here $N = 500$

## 4 Isotopic boundary between non-integrable domains.

Current section is devoted to the resistance at the isotopic boundary between the non-integrable domains. For modeling, we use two potentials that belong to different universality classes with respect to the bulk conductivity:

$\beta$-FPU potential:

$$V(u) = \frac{1}{2}u^2 + \frac{\beta}{4}u^4 \tag{40}$$

Periodic potential (chain of rotators):

$$V(u) = 1 - \cos u \tag{41}$$

### 4.1 $\beta$-FPU model

First, we check the asymptotic limit of very high mass mismatch, similar to Eq. (33), for the case of $\beta$-FPU model. The heat flux variation for mass-interface system is shown in Fig. 14 for very large $m_1$ and for very small friction $\gamma = 0.1$ at $T = 0.1 - 0.3$. In Fig. 14, the data can be collapsed into single equation for relatively large $m_1$, which is somewhat similar to the asymptotic equation (33), besides the temperature dependence.

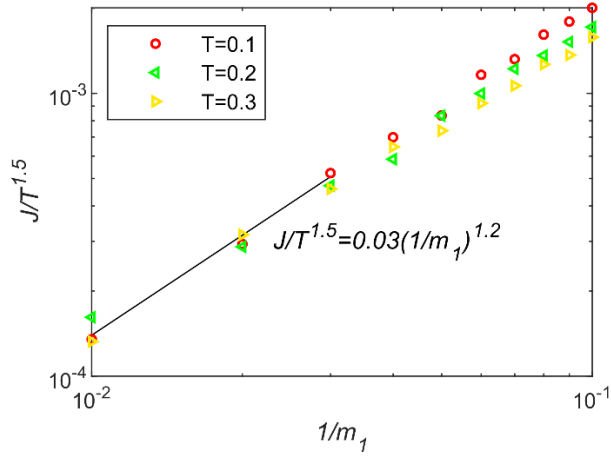

FIG. 14. (Color online) Heat flux variation with the chain temperature and $m_1$ for $\beta-FPU$ chain as $\gamma \ll 1, |m_1 - 1| \gg 1$. Here $N = 500, \beta = 0.1, \gamma = 0.1, \Delta = 0.1$

In Fig. 15, we present the size dependence of Kapitza resistance for $\beta$-FPU for various temperatures $T = 0.001 - 100$. At very low temperatures and for moderate chain size, the Kapitza resistance is very close to the size-independent value predicted for the linear chain. For the moderate chain length and high temperatures, one observes the decrease of the resistance according to the scaling law $R_k \sim N^{-0.3}$. For larger systems and intermediate temperatures, one observes a crossover. So, it is possible to conjecture that even for low temperatures one will observe substantial size dependence of the Kapitza resistance. For this sake, one should consider very long chains, beyond our numeric capabilities.

Besides, the Kapitza resistance is only meaningful when there exists a significant temperature drop at the interface. From the numerical studies, it is observed such significant temperature drop only exists at small to moderate chain lengths with sufficiently higher temperature in the non-integrable models. In Fig. 16, the details of the scaling of the Kapitza resistance are presented. It is commonly known from several numeric experiments [31] that $J \sim N^{-\frac{2}{3}}$ for the FPU model. For Kapitza resistance, there is a cross-over to the exponent -2/3 from a mass-mismatch-dependent exponent in the thermodynamic limit. For the small chain lengths, as shown in the inset of Fig. 16, the temperature drop $\Delta T$ is also proportional to mass-mismatch dependent exponent (not proportional to $N^{-1}$) resulting in more or less universal size dependence of the resistance.

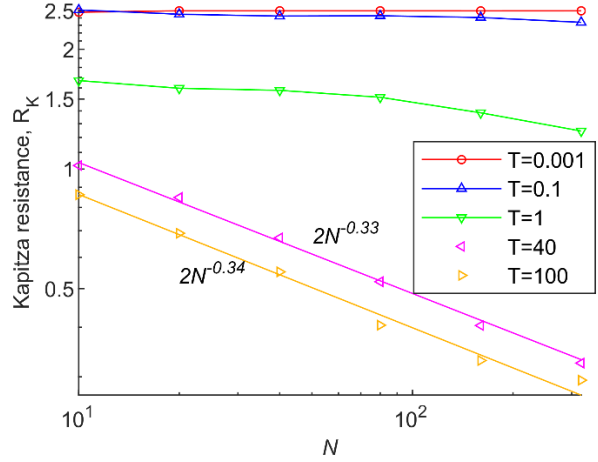

FIG. 15. (Color online) Numerical results of Kapitza resistance plotted by varying chain length $N$ for various temperatures in $\beta$-FPU model. At very low temperature and very small chain lengths, Kapitza resistance behavior is similar to linear model. $m_1 = 1.9, \gamma = 1, \Delta = 0.1$

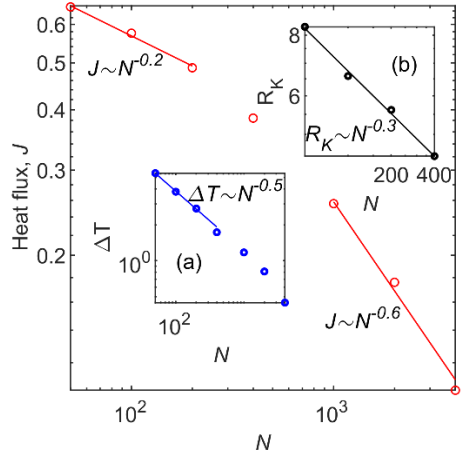

FIG. 16. (Color online) Scaling of heat flux with chain size is shown for FPU model. A significant crossover is observed. In the insets, the scaling of temperature drop (a) and Kapitza resistance (b) for small chain lengths are plotted. $T = 100, m_1 = 10, m_2 = 1, \Delta = 0.1$

In Fig. 17, the Kapitza resistance dependence on $m_1$ is shown. At $T = 0.01$ the plot resembles the harmonic case (cf. Fig. 3), although the square-root cusp is not observed clearly. At higher temperature $T = 3$ the nonlinearity fully reveals itself, and the plot clearly deviates from the harmonic case. In Fig. 18, the temperature dependence is shown. For very low temperatures, one encounters the linear limit (for given chain length). From intermediate temperature, a crossover is observed and the resistance monotonously decreases with the temperature. In addition to the dependence on the system size, the resistance also depends on the coupling friction.

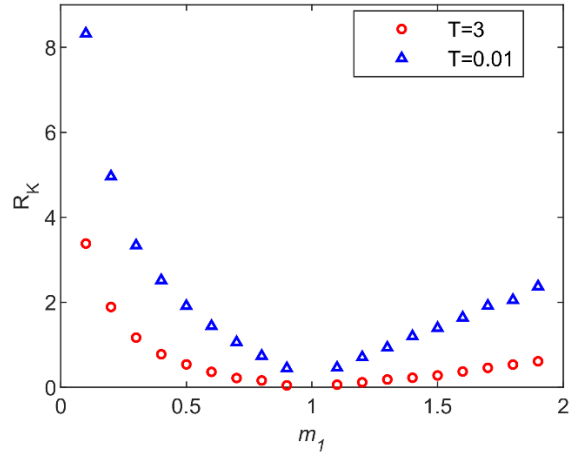

FIG. 17. (Color online) Kapitza resistance for the $\beta$-FPU model versus the mass mismatch $m_1$ ( $N=500, \gamma=1, \Delta=0.1$ ).

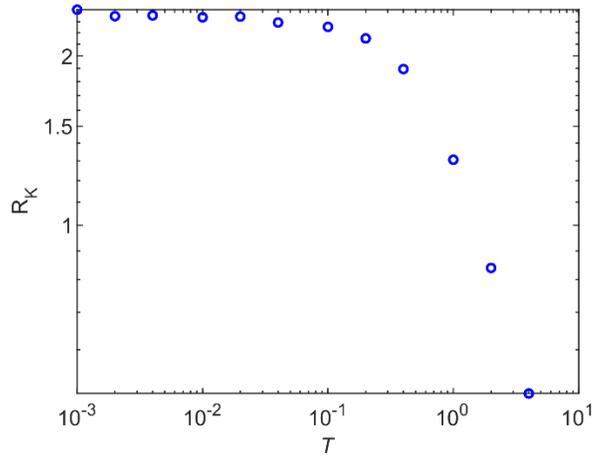

FIG. 18. (Color online) Kapitza resistance for the $\beta$-FPU model versus the average chain temperature $T$ ( $N=500, m_1=1.9, \gamma=1$ ).

### 4.2 Isotopic boundary between the rotator domains.

It is well-known that the rotator model exhibits normal bulk heat conductivity, since the ballistic transport phonons are locked by the rotobreathers [27]. The dependence of the Kapitza resistance on the system size is shown in Fig. 19. Similar to the FPU model, at very low temperatures and small system size the Kapitza resistance approaches the linear limit. Then, one observes an interesting behavior: for higher temperatures the size dependence saturates for moderate chain length, whereas for lower temperatures a crossover to (presumably) saturation is observed. Moreover, both $\Delta T$ (a) and $J$ (b) in the saturation regime scale normally, as $N^{-1}$ (see Fig. 20). In addition, the Kapitza resistance does not depend on the thermostat friction. The asymptotic mass dependence at $|m_1 - 1| \to 0$ conform to the linear case at low temperatures and deviates from it as temperature increases (see Fig. 21). In Fig. 22, the temperature dependence of Kapitza resistance is presented, and demonstrates the crossover from almost linear to substantially nonlinear behavior.

In the limit of the strong mass mismatch, one obtains the collapse of the heat flux, similarly to the β-FPU model (Fig. 23).

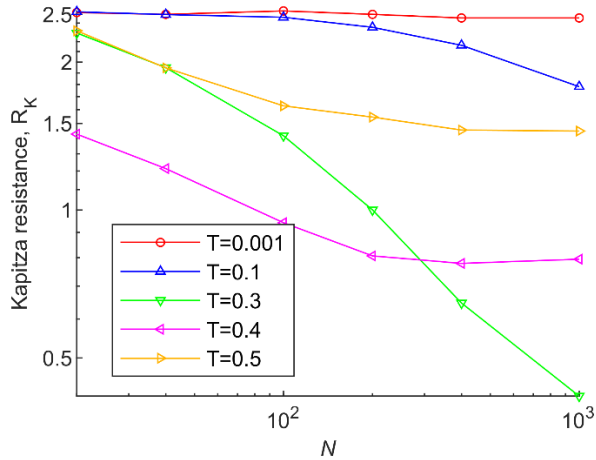

FIG. 19. (Color online) Kapitza resistance versus the chain length $N$ for various temperatures in rotator model. Here $m_1 = 1.9, \gamma = 1, \Delta = 0.1$. The lines are to guide the eye.

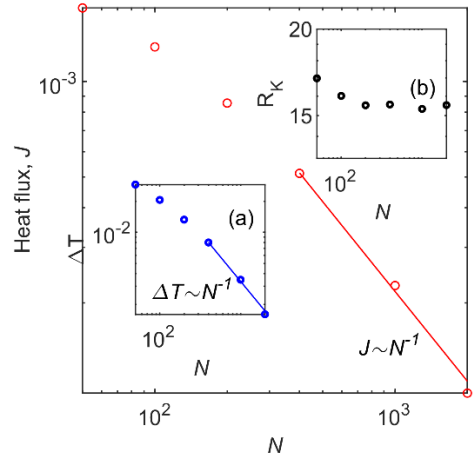

FIG. 20. (Color online) Scaling of heat flux with chain size is shown for chain of rotators. All the parameters showing normal behavior. Both $\Delta T$ (a) and $J$ (b) proportional to $1/N$, and $R_K$ does not depend on the chain size in the thermodynamic limit. $T_+ = 0.55, T_- = 0.45, m_1 = 10, m_2 = 1$

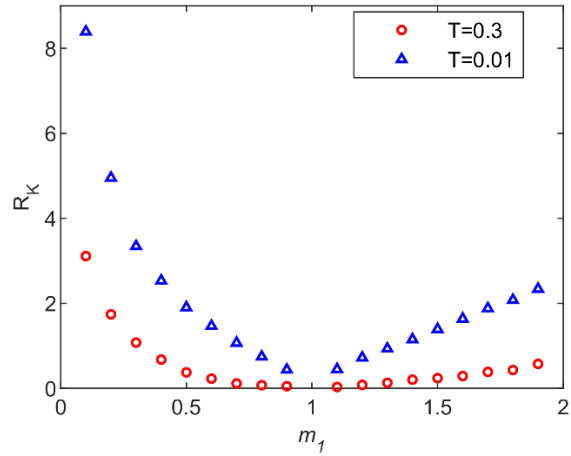

FIG. 21. (Color online) Mass-mismatch dependence of the Kapitza resistance for chain of rotators, $N = 500, \gamma = 1$.

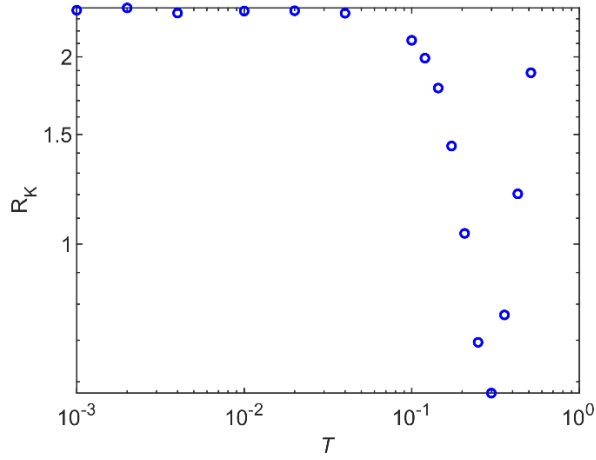

FIG. 22. (Color online) Temperature dependence of the Kapitza resistance for chain of rotators, $N = 500, m_1 = 1.9, \gamma = 1, \Delta = 0.1$.

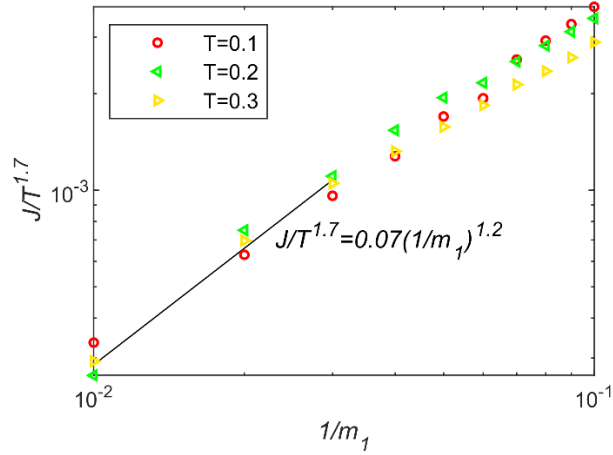

FIG. 23. (Color online) Heat flux dependence on $m_1$ for the chain of rotators at $\gamma \ll 1, |m_1 - 1| \gg 1$. Here $N = 500, \gamma = 0.1$

## 5 Conclusion

In the explored case of the domain boundary, one can derive the exact expression for the Kapitza resistance. Contrary to the case of isolated defect [21], generically the resistance is well-defined in the thermodynamic limit. However, in the linear case, as well as in the cases of Toda potential and colliding particles, the temperature drop and the heat flux do not depend on the system size. Therefore, the resistance is not a local property of the boundary, and should be still considered as anomalous. This anomalous non-locality also reveals itself in the dependence of the resistance on the thermostat characteristics. For the FPU chain, the anomaly is different – the resistance vanishes in the thermodynamic limit. The normal behavior with appropriate scaling behavior and locality is restored for the boundary between the rotator domains.

Contrary to the problem of the bulk conductivity, linear approximation is relevant for the nonlinear models (Toda, FPU, rotators) in the case of very small temperatures for the finite system size. As the system size grows for given temperature, one encounters the deviations from predictions of the linear theory – even weak nonlinearity reveals itself as the system size increases.

Consideration of the nonlinear integrable domains sheds light on the reason of thermal dependence on the Kapitza resistance. Indeed, for the chain of colliding particles one obtains the exact scaling $R_k \sim T^{-0.5}$. For the Toda lattice and can conjecture the crossover to a similar scaling for very high temperatures. For the FPU case, the final exponent may be different; this issue requires further exploration. For the chain of rotators, the temperature dependence is qualitatively different, presumably due to periodicity of the nearest-neighbor potential. For high mass mismatch $m_1 \gg 1$, one observes the data collapse of the form $J \sim T^\alpha m_1^\beta$. For exponent α, one has exact values $\alpha = 1$ for the linear system and $\alpha = 3/2$ for the colliding particles. For other explored models, the values are between or close to these limits. For the other exponent, one exactly derives $\beta = -3/2$. For the nonlinear models, this parameter exhibits substantial variability. This question also requires further exploration.

### Acknowledgment

The authors are very grateful to Israel Science Foundation (grant 1696/17) for financial support.

### References

[1]	N. Kürti, B. V. Rollin, and F. Simon, *Preliminary Experiments on Temperature Equilibria at Very Low Temperatures*, Physica **3**, 266 (1936).

[2]	W. H. Keesom and A. P. Keesom, *On the Heat Conductivity of Liquid Helium*, Physica **3**, 359 (1936).

[3]	P. L. Kapitza, *The Study of Heat Transfer in Helium II*, J. Phys. USSR **4**, 181 (1941).

[4]	G. L. Pollack, *Kapitza Resistance*, Rev. Mod. Phys. **41**, 48 (1969).

[5]     N. S. Snyder, *Heat Transport through Helium II: Kapitza Conductance*, Cryogenics **10**, 89 (1970).

[6]     L. J. Challis, *Kapitza Resistance and Acoustic Transmission across Boundaries at High Frequencies*, J. Phys. C Solid State Phys. **7**, 481 (1974).

[7]     E. T. Swartz and R. O. Pohl, *Thermal Boundary Resistance*, Rev. Mod. Phys. **61**, 605 (1989).

[8]     A. C. Anderson, J. I. Connolly, and J. C. Wheatley, *Thermal Boundary Resistance between Solids and Helium below 1°K*, Phys. Rev. **135**, A910 (1964).

[9]     E. T. Swartz and R. O. Pohl, *Thermal Resistance at Interfaces*, Appl. Phys. Lett. **51**, 2200 (1987).

[10]    I. M. Khalatnikov, *An Introduction To The Theory Of Superfluidity* (CRC Press, 2018).

[11]    W. A. Little, *The Transport of Heat between Dissimilar Solids at Low Temperatures*, Can. J. Phys. **37**, 334 (1959).

[12]    S. Simons, *On the Thermal Contact Resistance between Insulators*, J. Phys. C Solid State Phys. **7**, 4048 (1974).

[13]    R. E. Peterson and A. C. Anderson, *The Kapitza Thermal Boundary Resistance*, J. Low Temp. Phys. **11**, 639 (1973).

[14]    P. Erdös and S. B. Haley, *Low-Temperature Thermal Conductivity of Impure Insulators*, Phys. Rev. **184**, 951 (1969).

[15]    H. Budd and J. Vannimenus, *Thermal Boundary Resistance*, Phys. Rev. Lett. **26**, 1637 (1971).

[16]    W. M. Saslow, *Kapitza Conductance, Temperature Gradients, and Solutions to the Boltzmann Equation*, Phys. Rev. B **11**, 2544 (1975).

[17]    P. Taborek and D. L. Goodstein, *Diffuse Reflection of Phonons and the Anomalous Kapitza Resistance*, Phys. Rev. B **22**, 1550 (1980).

[18]    Ch. Steinbrüchel, *The Scattering of Phonons of Arbitrary Wavelength at a Solid-Solid Interface: Model Calculation and Applications*, Z. Für Phys. B Condens. Matter Quanta **24**, 293 (1976).

[19]    M. E. Lumpkin, W. M. Saslow, and W. M. Visscher, *One-Dimensional Kapitza Conductance: Comparison of the Phonon Mismatch Theory with Computer Experiments*, Phys. Rev. B **17**, 4295 (1978).

[20]    J. Paul and O. V. Gendelman, *Kapitza Resistance in Basic Chain Models with Isolated Defects*, Phys. Lett. A 126220 (2020).

[21]    O. V. Gendelman and J. Paul, *Kapitza Thermal Resistance in Linear and Nonlinear Chain Models: Isotopic Defect*, Phys. Rev. E **103**, 052113 (2021).

[22]     Z. Rieder, J. L. Lebowitz, and E. Lieb, *Properties of a Harmonic Crystal in a Stationary Nonequilibrium State*, J. Math. Phys. **8**, 1073 (1967).

[23]     G. Casati, *Energy Transport and the Fourier Heat Law in Classical Systems*, Found. Phys. **16**, 51 (1986).

[24]     T. Hatano, *Heat Conduction in the Diatomic Toda Lattice Revisited*, Phys. Rev. E - Stat. Phys. Plasmas Fluids Relat. Interdiscip. Top. **59**, R1 (1999).

[25]     Y. Du, H. Li, and L. P. Kadanoff, *Breakdown of Hydrodynamics in a One-Dimensional System of Inelastic Particles*, Phys. Rev. Lett. **74**, 1268 (1995).

[26]     S. Lepri, R. Livi, and A. Politi, *Heat Conduction in Chains of Nonlinear Oscillators*, Phys. Rev. Lett. **78**, 1896 (1997).

[27]     O. V. Gendelman and A. V. Savin, *Normal Heat Conductivity of the One-Dimensional Lattice with Periodic Potential of Nearest-Neighbor Interaction*, Phys. Rev. Lett. **84**, 2381 (2000).

[28]     A. V. Savin and O. V. Gendelman, *Heat Conduction in One-Dimensional Lattices with on-Site Potential*, Phys. Rev. E **67**, 041205 (2003).

[29]     J. W. S. Rayleigh, *Treatise on Sound Vol II* (London: Macmillan, 1878).

[30]     M. Toda, *Recent Advances in the Theory of Nonlinear Lattices*, in *Theory of Nonlinear Lattices*, edited by M. Toda (Springer, Berlin, Heidelberg, 1989).

[31]     S. Lepri, R. Livi, and A. Politi, *Thermal Conduction in Classical Low-Dimensional Lattices*, Phys. Rep. **377**, 1 (2003).

## APPENDIX. A: AVERAGING IN EXPRESSIONS (17)

$$\left\langle \frac{1}{|D|^2} \right\rangle_{\substack{N_1 \to \infty \\ N_2 \to \infty}} = \frac{1}{4\pi^2} \int_0^{2\pi} d\varphi_1 \int_0^{2\pi} d\varphi_2 \times$$

$$\times \frac{1}{\chi \left| \begin{array}{l} \cosh x e^{i(\varphi_1+\varphi_2+\theta_1)} c_{1,+} c_{2,+} + \sinh x c_{2,+} c_{1,-} e^{i(\varphi_2-\varphi_1+\theta_2)} - \\ -\sinh x c_{1,+} c_{2,-} e^{-i(\varphi_2-\varphi_1+\theta_2)} - \cosh x e^{-i(\varphi_1+\varphi_2+\theta_1)} c_{1,-} c_{2,-} \end{array} \right|^2 } \begin{array}{l} \zeta_1 = \varphi_1 + \frac{\theta_1-\theta_2}{2} \\ \zeta_2 = \varphi_2 + \frac{\theta_1+\theta_2}{2} \end{array} \quad \text{(A1)}$$

$$= \frac{1}{4\pi^2} \int_0^{2\pi} d\zeta_1 \int_0^{2\pi} d\zeta_2 \frac{1}{\chi \left( A e^{i\zeta_1} - B e^{-i\zeta_1} \right) \left( A^* e^{-i\zeta_1} - B^* e^{i\zeta_1} \right)} = \frac{1}{2\pi} \int_0^{2\pi} \frac{d\zeta_2}{\chi \left\| A \right|^2 - \left| B \right|^2 \right|}$$

$$A = \cosh x e^{i\zeta_2} c_{1,+} c_{2,+} - \sinh x c_{1,+} c_{2,-} e^{-i\zeta_2}, \quad B = \cosh x e^{-\zeta_2} c_{1,-} c_{2,-} - \sinh x c_{2,+} c_{1,-} e^{i\zeta_2}$$

$$\frac{1}{2\pi}\int_0^{2\pi}\frac{d\zeta_2}{\chi\left\||A|^2-|B|^2\right\|}\bigg|_{j=\exp(i\zeta_2)} = \frac{-i}{2\pi}\oint_{|j|=1}\frac{1}{\chi}\frac{dj}{j(Q-Pj-P^*/j)} = \frac{1}{\chi\sqrt{Q^2-4PP^*}}$$

$$Q = \left(|c_{1,+}|^2|c_{2,+}|^2 - |c_{1,-}|^2|c_{2,-}|^2\right)\cosh^2 x + \left(|c_{1,+}|^2|c_{2,-}|^2 - |c_{2,+}|^2|c_{1,-}|^2\right)\sinh^2 x$$

$$P = \left(|c_{1,+}|^2 - |c_{1,-}|^2\right)\sinh x \cosh x \left(c_{2,+}c_{2,-}^*\right)$$
(A2)

$$\left\langle\frac{\exp(i(2\varphi_2+\theta_1+\theta_2))}{|D|^2}\right\rangle_{\substack{N_1\to\infty\\N_2\to\infty}} = \frac{1}{4\pi^2}\int_0^{2\pi}d\zeta_1\int_0^{2\pi}d\zeta_2\frac{\exp(2i\zeta_2)}{\chi\left(Ae^{i\zeta_1}-Be^{-i\zeta_1}\right)\left(A^*e^{-i\zeta_1}-B^*e^{i\zeta_1}\right)} =$$

$$= \frac{1}{2\pi}\int_0^{2\pi}\frac{\exp(2i\zeta_2)d\zeta_2}{\chi\left\||A|^2-|B|^2\right\|} = \frac{-i}{2\pi}\oint_{|j|=1}\frac{1}{\chi}\frac{jdj}{j(Q-Pj-P^*/j)} =$$
(A3)

$$= \frac{Q-\sqrt{Q^2-4PP^*}}{2\chi P\sqrt{Q^2-4PP^*}}$$

$$\left\langle\frac{\exp(-i(2\varphi_2+\theta_1+\theta_2))}{|D|^2}\right\rangle_{\substack{N_1\to\infty\\N_2\to\infty}} = \frac{Q-\sqrt{Q^2-4PP^*}}{2\chi P^*\sqrt{Q^2-4PP^*}}$$
(A4)